\documentclass{article}

\usepackage{cite}
\usepackage{amsmath,amssymb,amsfonts}
\usepackage{algorithmic}
\usepackage{graphicx}
\usepackage{textcomp}
\usepackage{xcolor}
\usepackage{makecell}
\usepackage{multirow}
\usepackage{multicol}
\usepackage{geometry}
\usepackage{siunitx}
\usepackage{verbatim}

\usepackage{color}
\usepackage{latexsym}
\usepackage{amssymb}
\usepackage{gensymb}
\usepackage{subfigure} 
\usepackage{booktabs}
\usepackage{diagbox}
\usepackage{multirow}
\usepackage{array}
\usepackage{threeparttable}
\usepackage[utf8]{inputenc}
\usepackage{graphicx}

\usepackage{geometry}
\usepackage{amsmath}

\usepackage{algorithmic}

\usepackage{array}

\usepackage{authblk}
\usepackage{textcomp}
\usepackage{color}

\begin{document}

\title{NAND-like SOT-MRAM-based approximate storage for error-tolerant applications}

\author[1,$\dag$]{Min Wang}
\author[1,$\dag$]{Zhengyi Hou}
\author[1,$\dag$]{Chenyi Wang}
\author[1]{Zhengjie Yan}
\author[1]{Shixing Li}
\author[1]{Ao Du}
\author[1]{Wenlong Cai}
\author[1]{Jinhao Li}
\author[1]{Hongchao Zhang}
\author[1]{Kaihua Cao}
\author[1]{Kewen Shi}
\author[1]{Bi Wang}
\author[2]{Yuanfu Zhao}
\author[3]{Qingyi Xiang}
\author[1,*]{Zhaohao Wang}
\author[1]{Weisheng Zhao}

\affil[1]{School of Integrated Circuit Science and Engineering, Beihang University, Beijing, China}
\affil[2]{Beijing Microelectronics Technology Institute, Beijing, China}
\affil[3]{Huawei Technologies Co., Ltd, Shenzhen, China}
\affil[$\dag$]{authors contributed equally}
\affil[*]{corresponding author: zhaohao.wang@buaa.edu.cn}

\date{}
\maketitle

%



%



\begin{abstract}

We demonstrate approximate storage based on NAND-like spin-orbit torque (SOT) MRAM, through “device-modeling-architecture” explorations. We experimentally achieve down to 1E-5 level selectivity. Selectivity and low-power solutions are established by numerical calculation workflow. System-level power consumption is evaluated in the 512 KB last-level cache according to 5 quality levels. Error-tolerant applications, such as image processing, alleviate the demand for selectivity down to the 5E-2 level, leading to $54\%\sim 61\%$ energy-saving. Our proposal paves the novel and suitable path for high-density and low-power NAND-like SOT-MRAM.

\end{abstract}

\section{Introduction}
Emerging NAND-like SOT devices employ the structure of multiple bits arranged in the same SOT track\cite{cai2022selective,inokuchi2017improved,shi2021experimental}, which features area saving and high efficiency, but requires low-error-rate selective operation through modulation effect, e.g., spin transfer torque (STT) or voltage-controlled magnetic anisotropy (VCMA) effect. In-plane anisotropy (IMA) SOT-MTJ holds the field-free switching mechanism, the back-end of line (BEOL) compatibility, and radiation tolerance\cite{wang2021ionization,du2023electrical,honjo2019first,zhang2022integration}. However, pursuing the desired selectivity, e.g., 1E-6 level, requires high energy consumption and a strong modulation effect. Here, we propose and demonstrate that approximate storage relaxes the requirement of selectivity for NAND-like SOT devices and thus achieves significant energy-saving.

\section{Experiment and results}

We first fabricate 4\mbox{-}bit NAND\mbox{-}like SOT devices based on a 200\mbox{-}mm\mbox{-}wafer manufacturing platform with MgO(1.5 nm), noted as Process A (Fig. 1(a) and (b)) \cite{zhang2022integration}. Negative vertical voltage (\mbox{-}\(V_{M}\)) assists writing in IMA devices, as verified by micromagnetic simulation (Fig. 1(c)) and experiments (Fig. 2(b)\mbox{-}(d)) \cite{inokuchi2017improved}.  Statistics of TMR versus MTJ position and canted angles are listed in Fig. 2(a). Fig. 2(b)\mbox{-}(d) summarizes the switching probability (\(P_{sw}\)) and critical switching SOT voltage (\(V_{sw}\)) for various cases. Critical SOT current density is inferred as ~11.9 MA/cm$^{2}$@300 ns. An available window exists in the 1E-5 level selectivity by electrical measurement (Fig. 3(a)). Under the "ballooning", the selectivity of most devices is limited to the 1E-3 $\sim$ 1E-4 level, which fortunately does not limit their use as approximate storage. In addition, excessively high selectivity targets may lead to endurance problems (Fig. 3(b)). Fig. 4 shows the results of the two test methods, under the fixed \(V_{M}\) or fixed \(V_{TOP}\). Different control voltages should be designed for 4 MTJs to achieve better modulation, especially in \(\mu\)\(m\)-level tracks (Fig. 4(b)), due to the position\mbox{-}dependent IR drop along the SOT track. In addition, Process B prepared sub\mbox{-}100 nm devices (MgO$\sim$1.1 nm), and Fig. 5 shows the results at 5 ns pulses.

\section{Modeling and workflow}
The numerical calculation workflow (Fig. 6) is established by considering ({\romannumeral1}) equivalent temperature assumption, ({\romannumeral2}) temperature dependence of parameters, ({\romannumeral3}) magnetoresistance modeling, etc \cite{krizakova2021interplay,taniguchi2022magnetization}. The results for a single bit (Fig. 7(a)) are in good agreement with the experimental characteristics (Fig. 5). The contribution factors of STT and VCMA (\(c_{STT}\) and \(c_{VCMA}\)) can be further estimated. The dominant role gradually shifts from STT to VCMA with thicker MgO, as shown in Fig. 7(b).

\section{Solutions for selectivity}
In Fig. 8, energy evaluation is performed assuming a miniaturized track and 1E-2 level selectivity (i.e., approximate storage). As shown in Fig. 8(a), the required VCMA coefficient \(\xi \) decreases with thinner MgO, transitioning from the voltage-controlled to the current-controlled mechanism. The former has the advantage of ultra-low energy consumption (55.6 fJ@99$\%$\(P_{sw}\) and 1.7 nm MgO), noted as Solution A. Fig. 8(b) provides another possible way to balance energy consumption and requirement for \(\xi \). An increase in pulse width causes a decrease in controlling voltage, forming a local optimum at $\sim$ 20 ns pulse width (153 fJ@99$\%$\(P_{sw}\)), noted as Solution B.

\section{Image processing application}
As shown in Fig. 9(a), two 4-bit NAND-like SOT devices are utilized as the high-significance and low-significance bits (HSB and LSB) for 8-bit image data approximate storage.  Fig. 9(b) illustrates the workflow for satisfying different quality requirements by using the writing voltage and LSB as the quality knob. Fig. 9(c) shows the distribution of energy consumption in the 512 KB last-level cache. The device parameters for 5 quality levels (L1-L5) are listed in Table.1. The JPEG and Sobel processing outputs are shown in Fig. 10. As shown in Fig. 11(a), the results of root mean square error (RMSE) in L1-L4 are more acceptable compared with L5. Fig. 11(b) illustrates that $\sim$61$\%$ and $\sim$54$\%$ energy-saving are achieved from L1 to L4 for Solutions A and B, respectively. According to Table.2, the proposed approaches have a 40 to 323 times writing energy decrease than reported works \cite{monazzah2020cast}, with RMSE less than 30.

\section{Conclusion}
Our work demonstrates the BEOL compatibility, low power consumption, and adaptability of multi-terminal operation in arrays for NAND-like SOT devices. In image processing, NAND-like SOT devices serve as approximate storage with less requirement for selectivity (5E-2 level is enough) and $54\%\sim 61\%$ energy-saving from L1 to L4. Overall, our proposal opens up new application scenarios and opportunities for SOT-MRAMs.



\section*{Acknowledgements} \par 
This work was supported by the National Natural Science Foundation of China under Grant 62171013, 62271026, and 62104015, National Key Research and Development Program of China (Nos. 2021YFB3601303, 2021YFB3601304).


\bibliographystyle{ieeetr}
\bibliography{reference}

\begin{thebibliography}{10}

\bibitem{cai2022selective}
K.~Cai, S.~Van~Beek, S.~Rao, K.~Fan, M.~Gupta, V.~Nguyen, G.~Jayakumar,
  G.~Talmelli, S.~Couet, and G.~S. Kar, ``Selective operations of multi-pillar
  {SOT-MRAM} for high density and low power embedded memories,'' in {\em 2022
  IEEE Symposium on VLSI Technology and Circuits (VLSI Technology and
  Circuits)}, pp.~375--376, IEEE, 2022.

\bibitem{inokuchi2017improved}
T.~Inokuchi, H.~Yoda, Y.~Kato, M.~Shimizu, S.~Shirotori, N.~Shimomura,
  Y.~Kamiguchi, H.~Sugiyama, S.~Oikawa, K.~Ikegami, {\em et~al.}, ``Improved
  read disturb and write error rates in voltage-control spintronics memory
  ({V}o{CSM}) by controlling energy barrier height,'' {\em Applied Physics
  Letters}, vol.~110, no.~25, 2017.

\bibitem{shi2021experimental}
K.~Shi, W.~Cai, Y.~Zhuo, D.~Zhu, Y.~Huang, J.~Yin, K.~Cao, Z.~Wang, Z.~Guo,
  Z.~Wang, {\em et~al.}, ``Experimental demonstration of {NAND}-like
  spin-torque memory unit,'' {\em IEEE Electron Device Letters}, vol.~42,
  no.~4, pp.~513--516, 2021.

\bibitem{wang2021ionization}
B.~Wang, M.~Wang, H.~Zhang, Z.~Wang, Y.~Zhuo, X.~Ma, K.~Cao, L.~Wang, Y.~Zhao,
  T.~Wang, {\em et~al.}, ``Ionization and displacement damage on nanostructure
  of spin-orbit torque magnetic tunnel junction,'' {\em IEEE Transactions on
  Nuclear Science}, vol.~69, no.~1, pp.~43--49, 2021.

\bibitem{du2023electrical}
A.~Du, D.~Zhu, K.~Cao, Z.~Zhang, Z.~Guo, K.~Shi, D.~Xiong, R.~Xiao, W.~Cai,
  J.~Yin, {\em et~al.}, ``Electrical manipulation and detection of
  antiferromagnetism in magnetic tunnel junctions,'' {\em Nature Electronics},
  vol.~6, no.~6, pp.~425--433, 2023.

\bibitem{honjo2019first}
H.~Honjo, T.~Nguyen, T.~Watanabe, T.~Nasuno, C.~Zhang, T.~Tanigawa, S.~Miura,
  H.~Inoue, M.~Niwa, T.~Yoshiduka, {\em et~al.}, ``First demonstration of
  field-free {SOT-MRAM} with 0.35 ns write speed and 70 thermal stability under
  400°{C} thermal tolerance by canted {SOT} structure and its advanced
  patterning/{SOT} channel technology,'' in {\em 2019 IEEE International
  Electron Devices Meeting (IEDM)}, pp.~28--5, IEEE, 2019.

\bibitem{zhang2022integration}
H.~Zhang, X.~Ma, C.~Jiang, J.~Yin, S.~Lyu, S.~Lu, X.~Shang, B.~Man, C.~Zhang,
  D.~Li, {\em et~al.}, ``Integration of high-performance spin-orbit torque
  {MRAM} devices by 200-mm-wafer manufacturing platform,'' {\em Journal of
  Semiconductors}, vol.~43, no.~10, p.~102501, 2022.

\bibitem{krizakova2021interplay}
V.~Krizakova, E.~Grimaldi, K.~Garello, G.~Sala, S.~Couet, G.~S. Kar, and
  P.~Gambardella, ``Interplay of voltage control of magnetic anisotropy,
  spin-transfer torque, and heat in the spin-orbit-torque switching of
  three-terminal magnetic tunnel junctions,'' {\em Physical Review Applied},
  vol.~15, no.~5, p.~054055, 2021.

\bibitem{taniguchi2022magnetization}
T.~Taniguchi, S.~Isogami, Y.~Shiokawa, Y.~Ishitani, E.~Komura, T.~Sasaki,
  S.~Mitani, and M.~Hayashi, ``Magnetization switching probability in the
  dynamical switching regime driven by spin-transfer torque,'' {\em Physical
  Review B}, vol.~106, no.~10, p.~104431, 2022.

\bibitem{monazzah2020cast}
A.~M.~H. Monazzah, A.~M. Rahmani, A.~Miele, and N.~Dutt, ``Cast: Content-aware
  {STT-MRAM} cache write management for different levels of approximation,''
  {\em IEEE Transactions on Computer-Aided Design of Integrated Circuits and
  Systems}, vol.~39, no.~12, pp.~4385--4398, 2020.

\end{thebibliography}
%
%
%
%
%
%
%
%
%


%

\begin{figure}[htbp]
      \centering
            \includegraphics[width=0.5\linewidth]{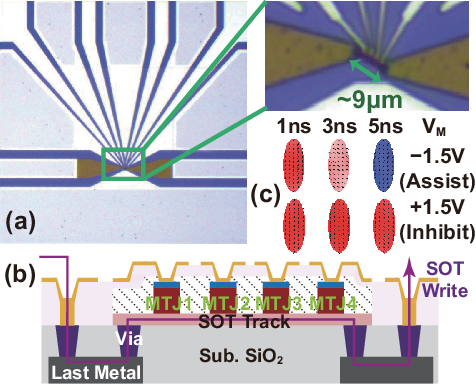}
      \caption{(a) Microscopy image and (b) schematic of the 4-bit NAND-like SOT device. (c) Micromagnetic simulation under various $V_{M}$.}
      \label{fig1}
\end{figure}

\begin{figure}[htbp]
      \centering
            \includegraphics[width=0.5\linewidth]{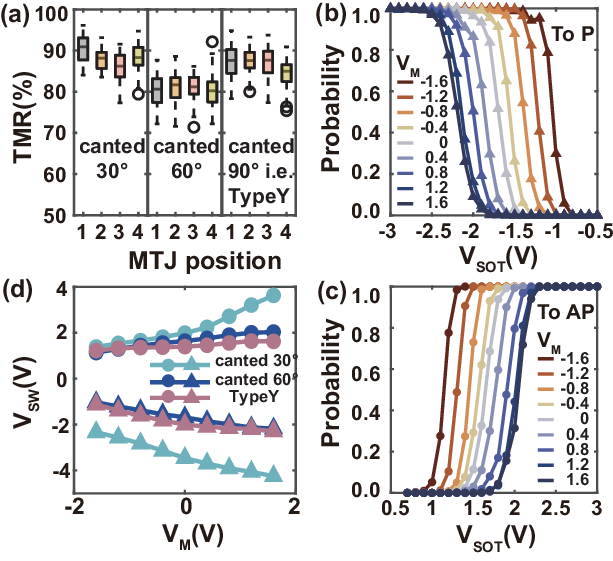}
      \caption{(a) TMR statistics (over 16 dies). Switching probability Psw versus $V_{M}$ during (b) P and (c) AP writing at 300 ns pulse. (d) Voltage modulation at various types of devices.}
      \label{fig2}
\end{figure}

\begin{figure}[htbp]
      \centering
            \includegraphics[width=0.3\linewidth]{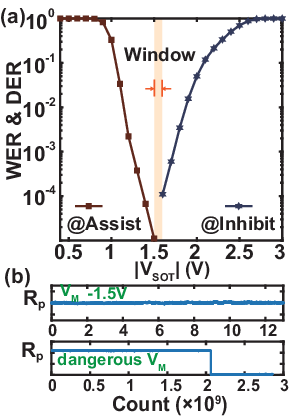}
      \caption{(a) 1E-5 level selectivity@500 ns. (b) Case of breakdown (MgO $\sim$1.3 nm)}
      \label{fig3}
\end{figure}

\begin{figure}[htbp]
      \centering
            \includegraphics[width=0.5\linewidth]{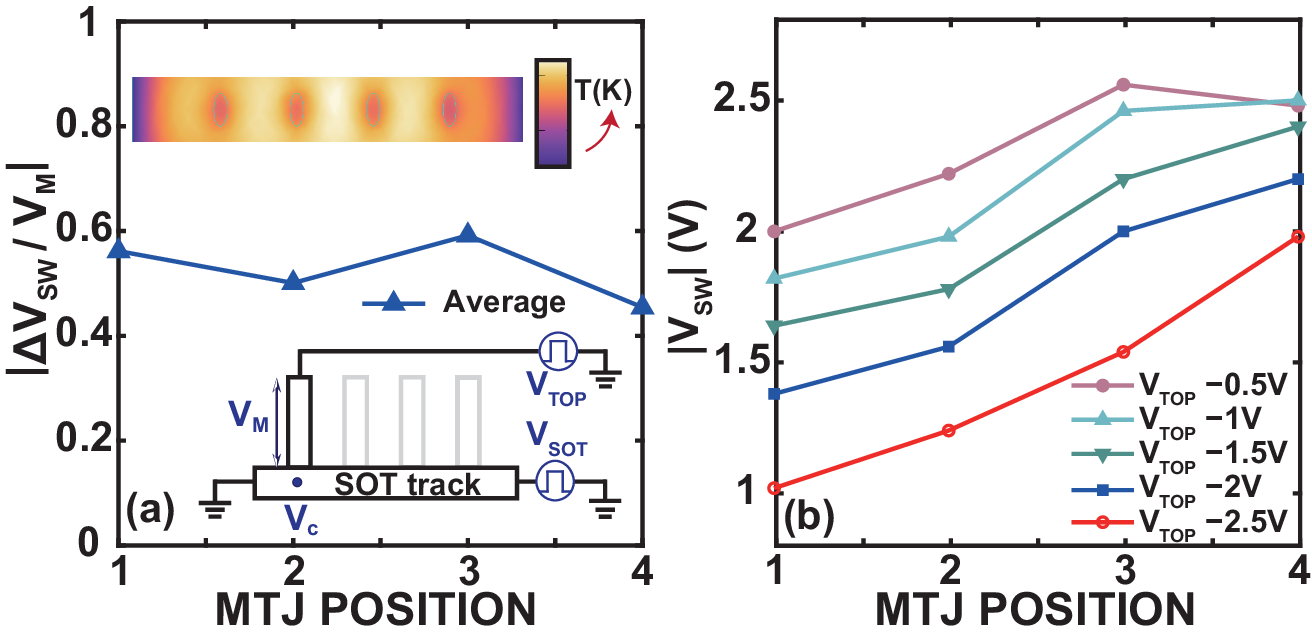}
      \caption{Experimental results for MTJ at various positions in 4-bit NAND-like devices. (a) Average $\left|  \Delta V_{sw}/V_{M} \right| $ under the same \(\pm\)$V_{M}$ in bipolar writing. Inset is the temperature distribution on the SOT track by Multiphysics simulation. (b) Dependence of $\left| V_{sw} \right|$   on the \(V_{TOP}\) and MTJ positions.}
      \label{fig4}
\end{figure}

\begin{figure}[htbp]
      \centering
            \includegraphics[width=0.3\linewidth]{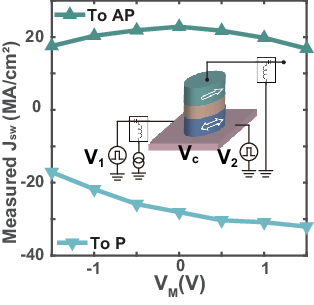}
      \caption{Experimental results of \(J_{sw}\)@5 ns as a function of $V_{M}$ in Process B, with \(J_{SOT}\) assumed as \(min\{J_{SOT}(V_{1}, V_{c}), J_{SOT}(V_{c}, V_{2})\}\)}
      \label{fig5}
\end{figure}

\begin{figure}[htbp]
      \centering
            \includegraphics[width=0.6\linewidth]{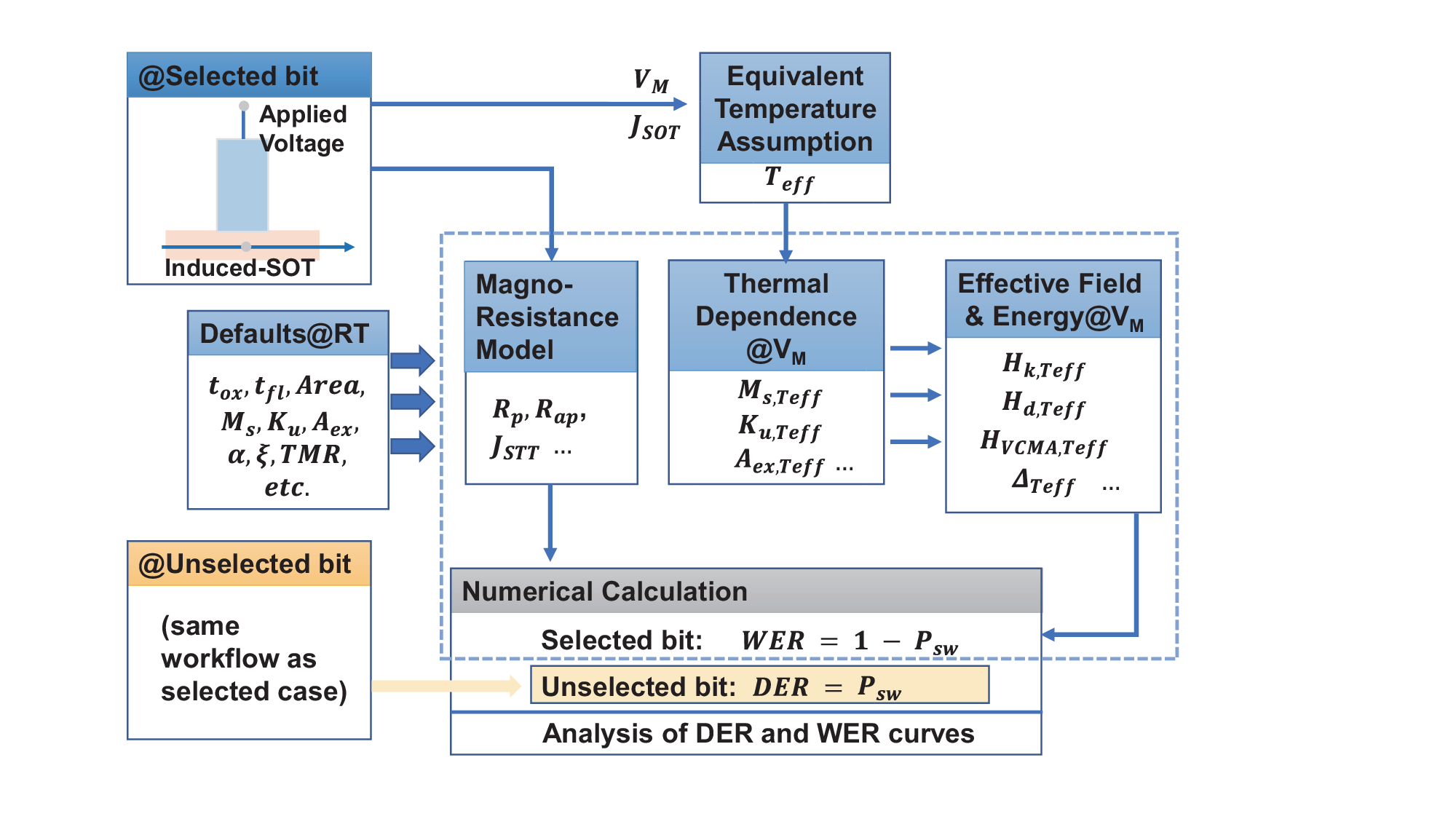}
      \caption{Numerical calculation workflow for modulation and selectivity analysis. }
      \label{fig6}
\end{figure}

\begin{figure}[htbp]
      \centering
            \includegraphics[width=0.3\linewidth]{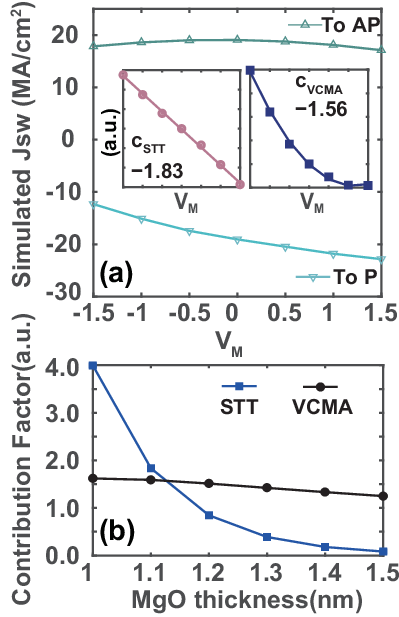}
      \caption{(a) \(J_{sw}\) as a function of $V_{M}$. (b) \(c_{STT} \) and \(c_{VCMA}\) in various MgO thicknesses. Normalized \(j_{sw}=1-(\pm c_{STT}V_{M} +c_{VCMA}V_{M}+{c_{T}V_{M}}^{2} )/ \left|J_{sw,0} \right| \).}
      \label{fig7}
\end{figure}

    \begin{figure}[htbp]
      \centering
            \includegraphics[width=0.4\linewidth]{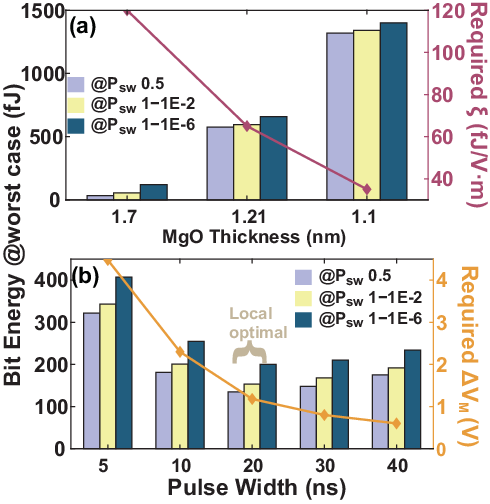}
      \caption{Energy consumption @ worst bit and (a) required \(\xi \) at different MgO thicknesses \(t_{ox}\) while fixing \(\Delta\)$V_{M}$ to 3 V (e.g., ±1.5 V). (b) required \(\Delta\)$V_{M}$ under various pulse widths while fixing \(\xi \) to 60 fJ/(V·m) and \(t_{ox}\) to 1.4 nm.}
      \label{fig8}
\end{figure}

    \begin{figure}[htbp]
	\centering
	\includegraphics[width=0.6\linewidth]{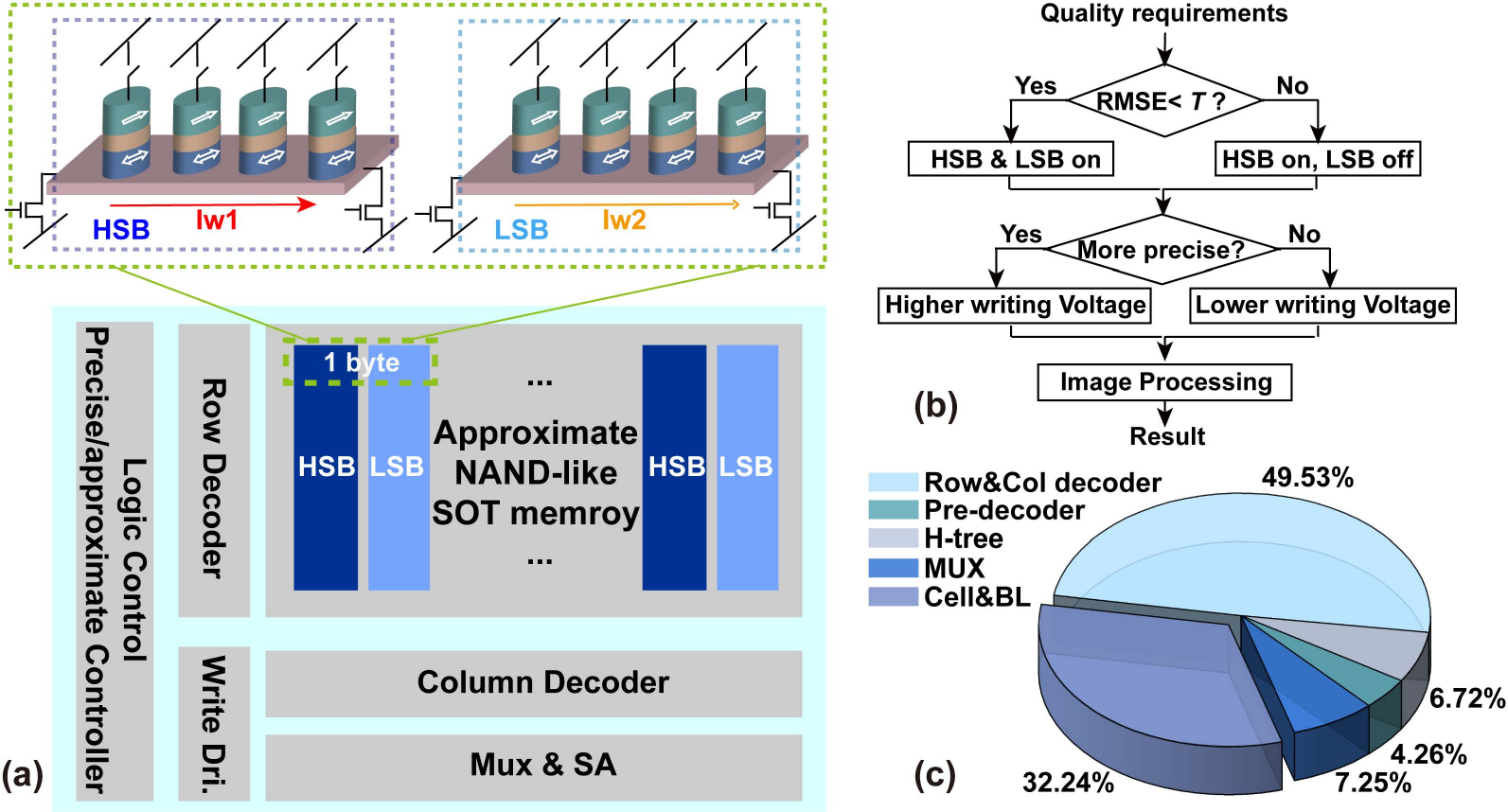}
	\caption{NAND-like SOT-based approximate storage concept. (a) The proposed structure and the 1-byte data cell with 4 bits HSB and 4 bits LSB. (b) The proposed workflow satisfies different quality requirements, where $T$ denotes the user-defined threshold. (c) Distribution of the energy consumption in a 512 KB approximate storage last-level cache.}
	\label{fig9}
\end{figure}

    \begin{figure}[htbp]
	\centering
	\includegraphics[width=0.9\linewidth]{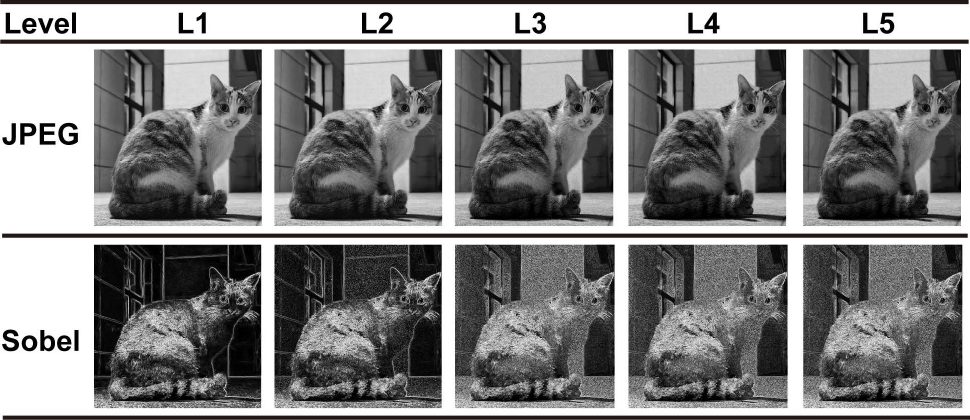}
	\caption{Output of JPEG and Sobel for 5 quality levels.}
	\label{fig10}
\end{figure}

   \begin{figure}[htbp]
      \centering
            \includegraphics[width=0.5\linewidth]{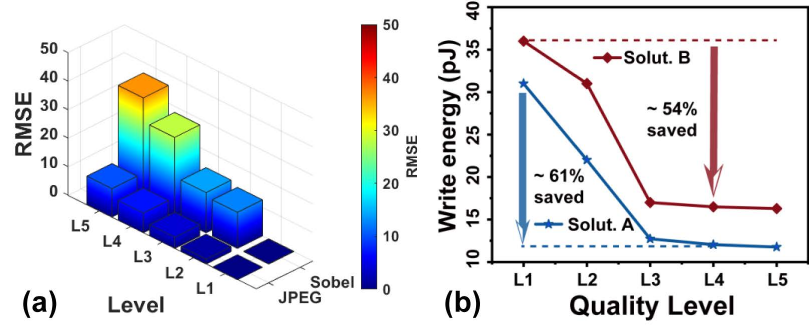}
      \caption{(a) RMSE results in different image processing applications, and (b) write dynamic energy consumption for an 8-byte cache line of two approaches.}
      \label{fig11}
\end{figure}

\begin{table}[h!]
    \begin{center}
    \caption{Quality-Energy map for NAND-like devices. $P_{sw}$ and write average energy per bit are obtained by calculation workflow.}
    \renewcommand{\arraystretch}{1.2}
    \setlength{\tabcolsep}{4mm}{
        \begin{tabular}{|cc|c|c|c|c|c|}
\hline
\multicolumn{2}{|c|}{Quality Level}                            & L1(baseline) & L2   & L3   & L4   & L5   \\ \hline
\multicolumn{1}{|c|}{\multirow{2}{*}{\(P_{sw}\)(\%)}}    & HSB        & 99.9999      & 99   & 99   & 95   & 90   \\ \cline{2-7} 
\multicolumn{1}{|c|}{}                            & LSB        & 99.9999      & 99   & 0    & 0    & 0    \\ \hline
\multicolumn{1}{|c|}{\multirow{2}{*}{\(E_{write}\)(fJ)}} & Solution A & 119          & 54.7 & 54.7 & 45.6 & 41.9 \\ \cline{2-7} 
\multicolumn{1}{|c|}{}                            & Solution B & 192          & 146  & 146  & 137  & 134  \\ \hline
        \end{tabular}}
    \end{center}
\end{table}

\begin{table}[h!]
	\caption{Performance comparison.}
	\begin{center}
		\begin{threeparttable} 
			\renewcommand{\arraystretch}{1.2}
			 
			\begin{tabular}{|c|c|c|cc|}
				\hline
				& QuARK\cite{monazzah2020cast}                     & CAST\cite{monazzah2020cast}                      & \multicolumn{2}{c|}{This work}               \\ \hline
				\multirow{2}{*}{Technology}                                           & \multirow{2}{*}{STT-MRAM} & \multirow{2}{*}{STT-MRAM} & \multicolumn{2}{c|}{NAND-like SOT-MRAM}      \\ \cline{4-5} 
				&                           &                           & \multicolumn{1}{c|}{Solution A} & Solution B \\ \hline
				Quality knob                                                     & \(V_{STT}\)                      & \(V_{STT}\)                     & \multicolumn{2}{c|}{\(V_{SOT}\) \& LSB }               \\ \hline
				\begin{tabular}[c]{@{}c@{}}\(E_{write}\)(pJ)\\ @worst case$^{\ast}$\end{tabular} & 672.25                    & 5325                      & \multicolumn{1}{c|}{12.04}      & 16.48      \\ \hline
				Energy-saving(\%)                                                & 47                        & 69                        & \multicolumn{1}{c|}{61}         & 54         \\ \hline
				RMSE @worst case                                                  & 31.93                     & 31.93                     & \multicolumn{2}{c|}{27.95}                   \\ \hline
			\end{tabular}
			\begin{tablenotes}  
				\footnotesize               
				\item[*] $ E_{write}$ is reported for an 8-byte cache line, L4 is acceptable worst case.         
			\end{tablenotes} 
		\end{threeparttable}  
	\end{center}
\end{table}

\end{document}